\pdfoutput=1
\documentclass[%
aps,prb,
showkeys,
amsmath,amssymb,
superscriptaddress,
twocolumn,
article]{revtex4-2}

\usepackage{graphicx}
\usepackage{lineno}
\usepackage{amstext}
\usepackage{xcolor}
\usepackage{algorithm2e}
\usepackage{multirow}
\usepackage{hyperref}
\hypersetup{
  colorlinks   = true, 
  urlcolor     = blue, 
  linkcolor    = blue, 
  citecolor   = red 
}

\begin{document}

\title{Single SiGe Quantum Dot Emission Deterministically Enhanced in a  High-Q Photonic Crystal Resonator}

\author{Thanavorn Poempool}
\email{thanavorn.poempool@jku.at}
\author{Johannes Aberl}
\affiliation{Institute of Semiconductor and Solid State Physics, Johannes Kepler University Linz, Altenbergerstra\ss e 69, 4040 Linz, Austria}
\author{Marco Clementi}
\altaffiliation[Current address: ]{Photonic Systems Laboratory (PHOSL), École Polytechnique Fédérale de Lausanne, Station 11, 1015 Lausanne, Switzerland}
\affiliation{Dipartimento di Fisica, Universit\`a di Pavia, Via Bassi 6, 27100 Pavia, Italy}
\author{Lukas Spindlberger}
\author{Lada Vuku\v{s}i\'c} 
\affiliation{Institute of Semiconductor and Solid State Physics, Johannes Kepler University Linz, Altenbergerstra\ss e 69, 4040 Linz, Austria}
\author{Matteo Galli} 
\author{Dario Gerace}
\affiliation{Dipartimento di Fisica, Universit\`a di Pavia, Via Bassi 6, 27100 Pavia, Italy}
\author{Frank Fournel}  
\author{Jean-Michel Hartmann}
\affiliation{University Grenoble Alpes, CEA, LETI, Grenoble, France}
\author{Friedrich Schäffler}
\author{Moritz Brehm}
\author{Thomas Fromherz}
\email{thomas.fromherz@jku.at}
\affiliation{Institute of Semiconductor and Solid State Physics, Johannes Kepler University Linz, Altenbergerstra\ss e 69, 4040 Linz, Austria}


\begin{abstract}
We report the resonantly enhanced radiative emission  from a single SiGe quantum dot (QD), which is deterministically embedded into a bichromatic photonic crystal resonator (PhCR) at the position of its largest modal electric field by a scalable method. By optimizing our molecular beam epitaxy (MBE) growth technique, we were able to reduce the amount of Ge within the whole resonator to obtain an absolute minimum of exactly one QD, accurately positioned by lithographic methods relative to the PhCR, and an otherwise flat, a few monolayer thin, Ge wetting layer (WL).  With this method, record quality (Q) factors for QD-loaded PhCRs up to $Q\sim 10^5$ are achieved. A comparison with control PhCRs on samples containing a WL but no QDs is presented, as well as a detailed analysis of the dependence of the resonator-coupled emission on temperature, excitation intensity, and emission decay after pulsed excitation. Our findings undoubtedly confirm a single QD in the center of the resonator as a potentially novel photon source in the telecom spectral range.

\end{abstract}


\maketitle
\section*{Introduction}

Driven by the rapid growth of the demand in data centers, silicon-based photonic integrated circuits (PICs) have recently witnessed rapid development. They have reached a level of maturity that allowed the implementation of foundry-type fabrication processes.\cite{2018_PotI_Chen,2021_JoLT_Siew} This swift progress was leveraged by utilizing extremely reproducible process steps, originally developed over decades for silicon complementary metal-oxide semiconductor (CMOS) integrated electric circuits to the emerging field of PICs. While for most building blocks of a PIC standard layout libraries are offered by numerous commercial foundries \cite{2021_JoLT_Siew}, an efficient, Si-based electrically driven light emitter is still missing, mostly due to silicon indirect bandgap. To overcome this hurdle, several methods of heterogeneous integration of III-V semiconductor lasers have been developed.\cite{2016_JoO_Thomson} Still, a monolithically integrated emitter would be desirable in terms of production costs and thermal stability. In this respect,  promising results on laser emission from  strained Ge\cite{2012_OE_CamachoAguilera,2015_OE_Koerner, Pilon2019}, GeSn \cite{2020_NP_Elbaz,2021_PR_Zhou} and SiGe QDs \cite{Grydlik2016} have been reported. 

Due to the versatile toolbox available, Si PICs are also highly attractive for possible applications in quantum photonics \cite{SilverstoneIJoSTiQE2016,2021_IJoSTiQE_Adcock,2021_AQT_Lu}. Complex quantum optical devices integrated on a Si chip have been demonstrated\cite{2019_NP_Wang}. Similar to classical integrated photonics, also in the quantum domain the indirect bandgaps of Si and Ge pose a major obstacle for the realization of deterministically emitting, monolithically integrated sources of quantum states of electromagnetic radiation. Hybrid integration of highly optimized III-V QDs single photon sources \cite{2017_NN_Senellart} on a Si based integrated quantum optical platform is thus a long standing goal, albeit the most advanced III-V QDs do not emit in the Si transparent wavelength region.\cite{2021_IJoSTiQE_Adcock} Also, defect centers in Si have gained recent attention as single photon sources in the telecom wavelength region.\cite{2021_PRL_Durand} However,  site controlled  integration of these defects into a PIC platform would be required for prospective scalability but has not been demonstrated, yet. 

Thus, despite the spatial separation of electrons and holes confined to SiGe QDs,\cite{2000_PRB_Schmidt} in addition to the restrictions imposed by the indirect band gaps of Si and Ge, single SiGe QDs as photon emitters remain an active area of research, mainly owing to their anticipated compatibility with CMOS technology,\cite{2009_PotI_Tsybeskov} to their natural emission in the telecom wavelength band and, crucially for CMOS compatibility and scalability, to the possibility of perfectly controlling their nucleation site (Ref. \cite{Grydlik2013} and references therein). Several routes to increase the photon emission efficiency of SiGe QDs have been developed, including Ge implantation during QD growth in an MBE reactor \cite{2016_NL_Grydlik,2021_PRB_MurphyArmando}, annealing and hydrogen diffusion for passivation of recombination centers\cite{2020_C_Spindlberger,2021_APL_Spindlberger} and radiative lifetime reduction by the Purcell effect \cite{1946PotA_Purcell} in photonic crystal resonators (PhCR)\cite{JannesariOE2014,2015_OE_Zeng,Schatzl2017}.

In this work, we concentrate on the enhancement of the emission efficiency by  site controlled integration of SiGe QDs into an advanced type of PhCR, providing a large Q-factor via simple design rules \cite{Alpeggiani2015}. Unlike in previous approaches, in which either more than one QDs were present in the photonic structure \cite{JannesariOE2014,2015_OE_Zeng} or all except one QD were removed from the photonic crystal resonator by etching \cite{Schatzl2017}, here we additionally employ a SiGe QD growth regime, in which \textit{a priori} exactly one QD is nucleating at the center of the PhCR. As a result, we achieve record Q-factor values for  PhCR cavities coupled to a QD emitter monolithically fabricated on a Si integrated optical platform. The  cavity coupled single QD emission occurs in the telecom wavelength region and is clearly observable up to room temperature.                
     
\section*{Experimental}
\subsection*{Photonic Crystal Cavities Design and Layout.}
In this work, we use a bichromatic-type PhCR \cite{Alpeggiani2015} to enhance the emission efficiency of SiGe QDs via the Purcell effect\cite{1946PotA_Purcell}. This effect becomes increasingly efficient for PhCRs with decreasing  mode volume ($V$) and increasing Q-factor ($Q$), ultimately scaling as the ratio $Q/V$.\cite{Andreani1999} In bichromatic PhCRs, both of these requisites can be effectively controlled by straightforward design rules.\cite{Alpeggiani2015,SimbulaAP2017}. A scanning electron micrograph (SEM) of a typical bichromatic PhCR used in this work is shown in Fig.\,\ref{fig:uncaped_perfect}(b). It consists of two main ingredients: a triangular lattice with hole radius $R$ and PhCR period $a$, and a line defect made of a row of holes with radius $r$ and period $a^{\prime}<a$. In Fig.\,\ref{fig:uncaped_perfect}(b), these four design elements are indicated by white labels. The ratio $\beta=\frac{a'}{a}$ most significantly determines the $Q$-factor of this type of PhCR. Since the line defect row has to end with holes centered at positions belonging to the underlying triangular lattice,  
$\beta$ can be expressed as $\beta=\frac{N}{N+1}$, where $N a$ is the center-to-center distance of outermost holes of the line defect. This distance  is then subdivided into $N+1$ periods of length $a'$, i.e $N a=(N+1)a'$. In simulations, $Q>10^9$ has been predicted  for an ideal Si-membrane PhCR with $\beta=0.96\ (N=24)$ \cite{Alpeggiani2015,SimbulaAP2017}. However, due to technological deviations from the ideal design, reported experimental values for such a PhCR were limited to $Q \sim 1.2\times 10^6$ in previous reports.\cite{SimbulaAP2017,Clementi2020}

On our chip, we fix $N=24$ and $R=100$\,nm for all PhCRs. The remaining design parameters were varied, resulting in a series of implemented PhCRs with lattice periods covering the range between 330\,nm and 430\,nm in 10\,nm steps. For each $a$, a subset of PhCRs with varying hole radii $r$ of the line defect with $r= 40,\,50,\,60,\,70$\,nm was fabricated. The vertical outcoupling efficiency was controlled by a so-called far-field optimization (ff) structure based on the second-order Bragg grating effect \cite{Tran2009, Portalupi2010}. For this purpose, the radius of each second hole around the main cavity [marked in yellow in Fig.\,\ref{fig:uncaped_perfect} (b)] was slightly increased by $\Delta r= 0$, 3 or 6\,nm, labeled ff0, ff3, and ff6 in the following. Thus, in total, 132 bichromatic PhCRs, each with a different combination of $a$, $r$ and $\Delta r$ were implemented on one chip. However, since  strong optical signals could be observed in our setups already with ff3 PhCRs, including ff6 PhCRs in the experiments reported here turned out to be not required.      

\subsection*{Sample Growth and Fabrication.}
In order to maximize the radiative transition rate between electrons and holes confined into a QD via the Purcell effect, a PhCR has to be fabricated in exact registry with the QD, such that the latter occupies the position of maximum electric field amplitude of the PhCR's electromagnetic ground mode. To achieve this alignment, we follow the route described in Ref.~\cite{Schatzl2017}. In brief, a custom-made  silicon-on-insulator (SOI) substrate with reduced background emission in the wavelength range relevant for this work (1.3-1.6\,$\mu$m) and a  70\,nm Si device layer on top of a 2\,$\mu$m thick buried oxide layer (BOX) was employed as an initial substrate for further growth and PhCR processing. Background emission as described above has recently been observed for commercial SOI substrates even at room temperature\cite{2012_NJP_Hauke,Shakoor2012} and was ascribed to remnants of the hydrogen implantation for the SOI smart-cut process\cite{1997_MSaE_Maleville}.  By a bonding/splitting/thinning/annealing process sequence\cite{2001_ECSProc_Moriceau} any potential optical active defects were removed in a similar way as reported in Ref.~\cite{2012_NJP_Hauke}. In addition, thinning of the device layer to just 70\,nm allowed us to vertically center the Ge QDs in the designed slab thickness of 220\,nm.

\begin{figure*}
\includegraphics[width=12cm]{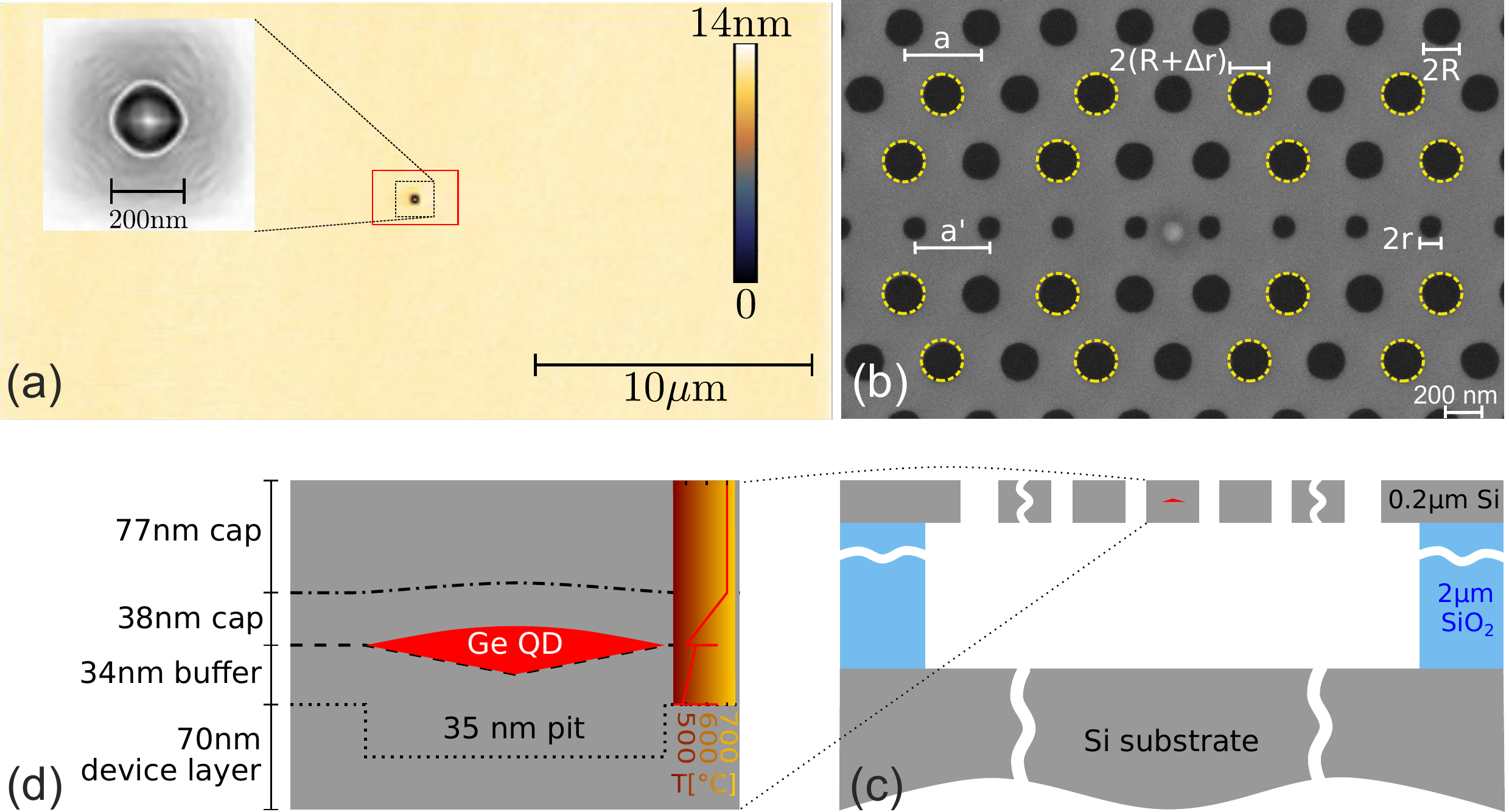}
\caption{(a) $30\times15\,\mu$m$^2$ AFM image of an uncapped sample with a single SiGe QD nucleated at the position predefined by a pit on a SOI substrate. No QDs outside the pits are observed. The inset shows a QD in a pit on larger scale. The red rectangular frame indicates the scan area shown in (b). (b) SEM image of an uncapped sample after aligned processing of the bichromatic resonator. The depicted resonator and QD are in virtually perfect registry in horizontal direction and in vertical direction with a deviation of less than 25\,nm.  Yellow dashed circles indicate air holes with radius enlarged by $\Delta r = +3$\,nm with respect to the unmarked ones outside the line defect. They form a second-order grating and allow precise control of the outcoupling of radiation and its far-field distribution. The relevant design parameters of the PhCR as described in the text are indicated by white lines and labels. (c) Cross-sectional sketch along the center of a bichromatic PhCR's line defect. The SiO$_2$ below the resonator was removed during HF etching through the PhCR holes as described in the text. (d) Cross-sectional sketch of the PhCR membrane's center. The QD indicated in red is vertically centered with respect to the membrane. Broken lines indicate approximate shapes of the sample surface after pit etch (dotted line), buffer layer growth (dashed line), and temperature-graded cap layer growth (dash-dotted line). The inset indicates the temperature profile employed during layer growth as described in the text. }
\label{fig:uncaped_perfect}
\end{figure*}

Samples were patterned by electron beam lithography (Methods). A three-mask-level process  
 allows the aligned exposure of the QD positioning mask and the PhCR mask with respect to alignment marks defined in the first lithographic step. The alignment marks were plasma-etched (Methods) into the SOI substrate to a depth of $\sim 110$\,nm. In the second step, a \textit{single} pit (SP) per PhCR was exposed at a defined position with respect to the alignment marks. This is distinctly different from our previous work \cite{JannesariOE2014,Schatzl2017}, where a pit \textit{array} was defined with all but the center pit in registry with  air holes of the PhCR to be defined in the final lithographic layer. The improvements resulting from this modification will be discussed below.
These pits were transferred into the SOI substrate by plasma etching at $-90^\circ$C as described in the Methods section. The etch depth was adjusted to 35 $\pm$ 5 nm as confirmed by atomic force microscopy (AFM). The cross-section of the device layer after pit-etching is schematically indicated by the dotted line in Fig.~\ref{fig:uncaped_perfect}(d).   At this point, the SP pre-patterned SOI substrate was ready for the QD growth process.

The QD growth procedure started with cleaning, degassing and buffer layer growth steps that are standard for MBE growth of SiGe layers. They are detailed in the Methods section. A sketch of the surface profile after the buffer deposition is shown in Fig.~\ref{fig:uncaped_perfect}(d) by the dashed line. Next, $\sim4$ monolayers (ML) Ge corresponding  to $\sim 5.69$\,\AA \  were deposited at a growth rate of 0.04\,\AA/s while the substrate temperature was maintained constant at 650$^\circ$C. Since the total Ge coverage stays below the critical one for spontaneous Ge QD formation on flat substrate areas (4.2-4.9\,ML at  650$^\circ$C), a two dimensional Ge WL without QDs forms there \cite{Grydlik2013}. On the other hand, QD formation in the pit sets in already at a Ge coverage much smaller than 4\,MLs. In addition, at this Ge coverage and the employed growth temperature, the Ge surface diffusion length is much larger than $30\,\mu$m \cite{GrydlikPRB2013} so that sufficient Ge for the formation of a QD in a pit reaches one of these sparse pits \cite{Grydlik2013}. We want to emphasize that a pit spacing of 100\,$\mu$m , as determined in this work by the center-to-center PhCR distance, is the largest for  perfectly site-controlled  QD formation  ever reported. It exceeds previously reported ones by at least  a factor $\sim30$ \cite{Grydlik2013,Grydlik2015,2015_OE_Zeng}. As predicted in Refs.\,\cite{Grydlik2013,GrydlikPRB2013}, no fundamental limit on the distance of site-controlled SiGe QDs seems to exist under suitable growth condition. Finally, samples intended for QD emission experiments were capped by a 115\,nm thick Si layer as described in the Methods section. As a result, the total thickness of grown layers plus original device layer is $\sim$220\,nm,  with a QD positioned in its center. Note that this total thickness is widely accepted as standard for the SOI integrated optical platform \cite{2016_JoO_Thomson}, allowing for a straightforward combination of the QD layer stack with device layouts readily available in PIC design libraries. Samples intended for structural characterization by surface scanning methods were grown without the Si cap. 

After MBE growth, the samples were prepared for the PhCR mask layer in the same way as for the previous lithographic steps. Due to the large etch depth of the alignment marks, a large fraction of them still shows sufficient contrast for automatized alignment by the e-beam system after the growth of $\sim 150$\,nm Si on top of them. With good alignment marks in place,  individual PhCRs could be  aligned with $\sim25$\,nm accuracy relative to the SQD positions [see Fig.~\ref{fig:uncaped_perfect}(b)]. Once again, the PhCRs were transferred into the sample by dry etching as described above. The etch time was adjusted for etching all the way down to the BOX. Compared to our previous work \cite{JannesariOE2014,Schatzl2017}, here, no QDs are present at the positions of the PhCRs' air holes, i.e. in the etch volume.  Therefore, a more homogeneous etch rate can be expected, resulting in a tighter etch time control and steeper air hole sidewalls.

Finally, free-standing PhCR membranes as sketched in Fig.~\ref{fig:uncaped_perfect}(c) are implemented by lateral under-etching through the PhCR air holes during a $\sim 24$\,minutes wet chemical etch in 10\%-HF. After this etch, the  2$\mu$m thick BOX below the PhCRs is removed. Identically processed, uncapped samples were used to confirm the alignment accuracy by scanning electron microscopy (SEM, LEO Supra 35 from Zeiss).

In addition to SQD samples, reference samples grown on unpatterned SOI substrates were investigated. These reference samples contain 
the same layer sequence as the SQD samples, including a cap layer. Each layer was grown under nominally identical conditions as its counterpart in the SQD samples. As discussed above, in the absence of pits the deposited Ge forms a homogeneous two-dimensional layer of constant thickness without QDs. In the following, we refer to this type of reference sample as a wetting layer (WL) sample. To fabricate reference PhCRs on the WL  samples, only the third level electron beam lithography step described above and the HF etch step are required.      
 
\begin{table}
\caption{List of SQD and reference samples onto which sets of PhCRs with systematically varying design parameters as described in the text were processed.} 
\label{tab:listofstructures}

\begin{tabular}{c  c c  c}
\hline \hline
sample ID &  emitter & discussed PhCRs  \\ \hline
 \multirow{2}{*}{QD}  & \multirow{2}{*}{SiGe SQD, Ge WL}  & a360 r70 ff3    \\ 
                     &                             & a360 r70 ff0 \\ \hline

\multirow{2}{*}{WL} & \multirow{2}{*}{Ge WL}  & a360 r70 ff3    \\
                  &                            & a350 r70 ff3 \\ 

\hline \hline
\end{tabular}
\end{table}


\section*{Results and Discussion}
\subsection*{Post-Processing Characterization} 
Figure\,\ref{fig:uncaped_perfect}(a) shows the morphology of a SQD that is nucleated on a SP pre-patterned substrate. The SQD is surrounded by a flat area free from randomly nucleated QDs as shown by a $30\times15\ \mu$m$^2$  AFM image in Fig.~\ref{fig:uncaped_perfect}(a). 
This perfect site control for QD nucleation is achieved by adjusting the  Ge surface coverage to stay below the supersaturation regime of $\sim 4.5$\,ML \cite{Brehm2009}. The area shown in Fig.~\ref{fig:uncaped_perfect}(b) corresponds to the area within the red rectangle in Fig.~\ref{fig:uncaped_perfect}(a). Figure\,\ref{fig:uncaped_perfect}(b) shows a SEM image of a bichromatic PhCR fabricated to be accurately aligned  with the QD, in order to maximize the dipole interaction of the QD with the electric field of the PhCR ground mode. 
The image was taken on an uncapped control sample, for which the QD is not hidden by the capping layer that is required for optically active QDs. 
The PhCR and QD are in virtually perfect registry in horizontal direction and in vertical direction with deviation of less than 25\,nm. The remaining misalignment is much smaller than typical QD dimensions.  It indicates limitations imposed by the employed fabrication technology in a similar range, as previously reported \cite{2015_OE_Zeng}. The yellow dashed circles in Fig.\,\ref{fig:uncaped_perfect} (b) mark photonic crystal  air holes with radii slightly modified by $\Delta$r. By these holes we are able to control the trade-off between highest $Q$-factor and  ff-coupling efficiency. 

We want to emphasize, that the QD-free area covers the whole $100\times100\,\mu\text{m}^2$ unit cell of the pit array outside the pit. 
Therefore, no secondary QD emitters are present to contribute to the resonant emission (RE) from PhCRs. Such a configuration is  in contrast to the work of Zeng et al. \cite{2015_OE_Zeng}, where PhCRs overlapping with more than one QD separated by 2\,$\mu$m QD are reported.

\begin{figure*}
\centering\includegraphics[width=11cm]{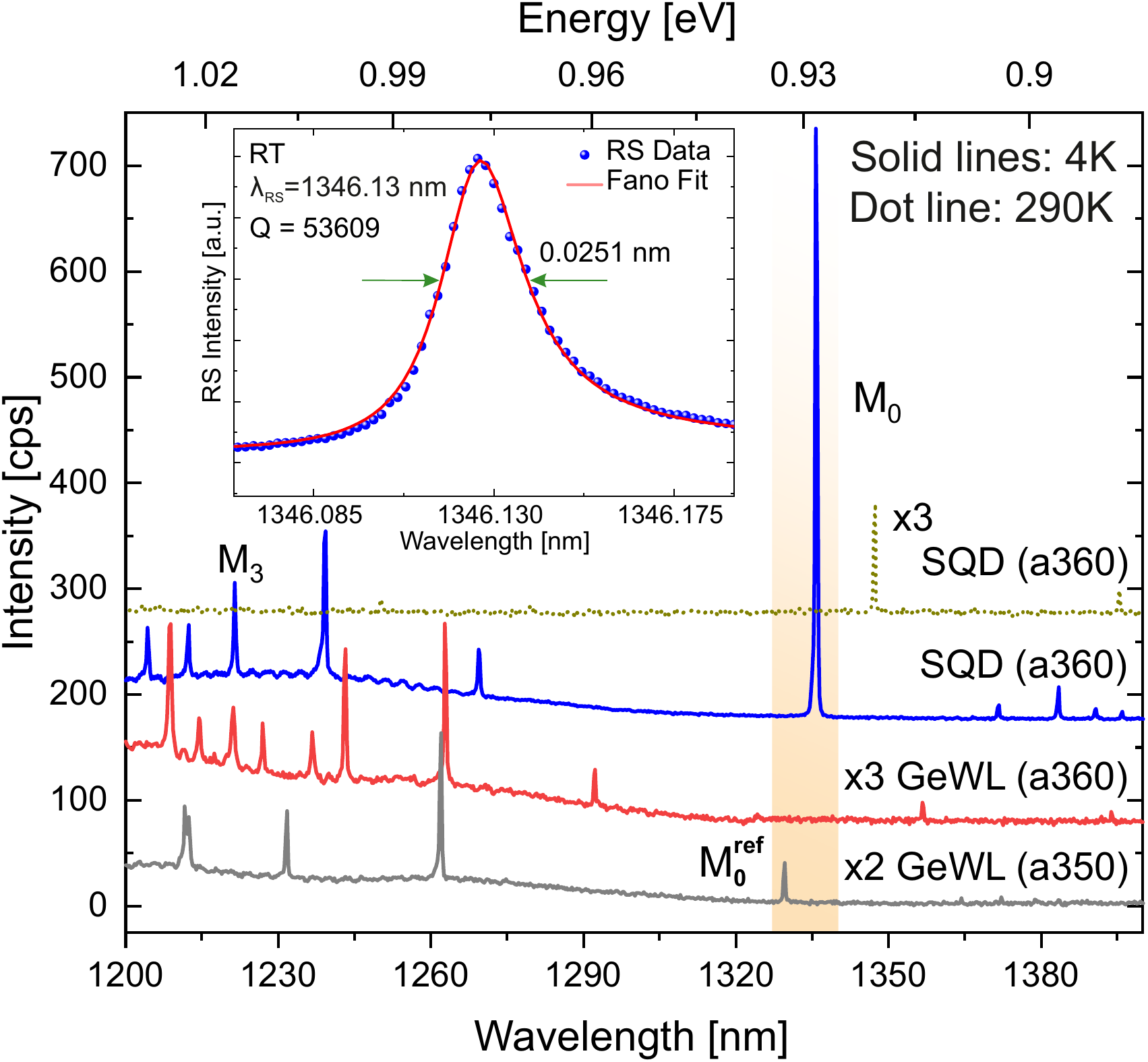}
\caption{Resonantly enhanced PL emission spectrum of the SQD centered in a N24 bichromatic PhCR  with parameters a360 r70 ff3 (blue line) measured at 4K. $M_{0}$ indicates the fundamental mode of the PhCR and $M_{3}$ indicates the third excited mode in resonance with optical wetting layer transitions. For comparison, the emission spectrum of this PhCR  is shown for a temperature of 290\,K by the dotted line on a scale expanded by a factor 3 (labeled $\times3$). The $\sim$11\,nm red-shift of the $M_0$-resonance wavelength is due to the increase of the Si refractive index with temperature. The red line shows the emissions from the Ge WL reference sample measured for a PhCR with nominally the same parameters as the one containing the QD (blue line). The intensity scale for this spectrum is three-fold expanded (indicated by label $\times3$). The gray line shows the emissions from the same Ge WL reference sample coupled to a cavity with different parameters (a350 r70 ff3), expanded by a factor 2 (labeled $\times2$). For this cavity layout, the spectral features, in particular the ground mode labelled $M_0^\text{ref}$, are more similar to the blue spectrum as compared to the spectrum shown in red. The shaded wavelength range indicates the  width of the ensemble QD emission observed before PhCR fabrication (see Fig.\,S1 in Supplementary Material). The inset shows the resonant scattering result obtained at room temperature for the SQD - PhCR system with PL emission shown in blue. A Q-factor of 53605 is determined. The resonance frequency coincides with the PL emission peak measured at 290\,K. 
}
\label{fig:comp3samples}
\end{figure*}

\subsection*{Optical Characterization} 
The emission properties of the cavity coupled QDs were studied by micro-photoluminescence ($\mu$-PL) spectroscopy (Methods). In Fig.~\ref{fig:comp3samples}, we compare the emission of a SQD at the center of a $N24$ bichromatic PhCR with results from reference PhCRs fabricated on a WL sample. 
As a typical example, the emission spectrum of a QD in a PhCR with $a=360$\,nm (a360) $r=70$\,nm (r70)
and second order out-coupling grating formed by air holes with nominal radius of 103\,nm ($\Delta r = +3$\,nm, ff3) is shown in Fig.\,\ref{fig:comp3samples} by the blue line. The emission spectrum is dominated by sharp resonances corresponding to the modes of the photonic structure. At 4\,K, the dominant peak (labelled $M_0$) is observed at $\sim 1336$\,nm wavelength ($\sim 0.9281$\,eV photon energy). As shown by the dotted line in Fig.\,\ref{fig:comp3samples}, increasing the temperature to 290\,K results in a $\sim 11$\,nm red-shift of the $M_0$-resonance as a consequence of the Si refractive index's increase with temperature. We attribute the $M_0$ resonance in Fig.\,\ref{fig:comp3samples} to the emission of the SQD amplified in weak coupling regime by the ground mode of the PhCR according to the Purcell effect \cite{1946PotA_Purcell}. As shown in the Supplementary Material (Fig.\,S2), the mode $M_0$ can be tuned in the wavelength range between $\sim1304$\,nm and $\sim1339$\,nm by selecting a PhCR period in the range 350\,nm  $\leq a \leq$  370\,nm. For all these different PhCR lattice constants, a strong SQD emission coupled to mode $M_0$ is observed, indicating a broad emission spectrum of a single SiGe QD, in agreement with the the results reported in Ref. \cite{Grydlik2015}. 

To corroborate our conjecture, we note that in the red and gray spectra of Fig.\,\ref{fig:comp3samples} the intense cavity-coupled SQD emission is absent and only faint lines are observed instead in the corresponding spectral region. These two spectra were measured on the WL sample, that contains no QDs. The red spectrum was observed for a PhCR nominally identical to the one corresponding to the spectrum for the SQD sample (blue line in Fig.\,\ref{fig:comp3samples}). However, the spectral features of both resonators show distinct differences too large to be attributed to limitations of chip-to-chip PhCR reproducibility. Instead, we rather ascribed to them to unequal Ge distributions within the PhCR for the SQD and the WL samples, and thus, to different effective refractive indices for these two nominally identical PhCRs. 
This effect can be compensated by a slight spectral tuning of the resonance wavelengths via the PhCR lattice constant $a$. As shown by the gray line in Fig.\,\ref{fig:comp3samples}, reducing the PhCR period $a$ from 360\,nm to 350\,nm results in resonance wavelengths resembling, up to a small blue-shift, those observed for the SQD sample. Most importantly, also for the spectrally tuned resonator on the WL sample, only a small peak (labelled $M_0^\text{ref}$ in Fig.\,\ref{fig:comp3samples} ) is observed  within the spectral range of resonance-coupled SQD emission shown in  Fig. S2 of the Supplementary Material. Therefore, the strong resonance $M_0$ is clearly due to the presence of a QD in the PhCR. More evidence for assigning the emission coupled to the $M_0$ mode to the SQD is obtained from its temperature- and excitation-power-dependence as discussed in  upcoming paragraphs.  

The origin of the faint resonances that are observed also in the absence of QDs in the spectral wavelength range above 1300\,nm  is attributed to residual WL emisson and/or emission associated with defects either in the SOI substrate,\cite{LoSavioAPL2011} or in the MBE grown Si. 
The main contribution of the WL sets in at wavelengths shorter than 1290\,nm, indicated by the broad background emission originating from photonic crystal regions outside the actual PhCR with superimposed resonances from WL regions inside the PhCR  shown in Fig.\,\ref{fig:comp3samples}. In this broad spectral region, for all PhCRs compared in Fig.\,\ref{fig:comp3samples} rather similar mode intensities are observed (note the different scaling factors for the blue, red, and gray spectra). This is especially true for the two PhCRs with different lattice constants $a$ on the WL sample, where the residual intensity differences are tentatively assigned to different Q-factors and outcoupling efficiencies of the various modes. For the SQD sample, the WL is thinner, since Ge from the WL is consumed during QD formation \cite{GrydlikPRB2013,Schuster2021}, thus, more dissimilarity to the WL-only samples can be expected.

\begin{figure}[t]
\includegraphics[width=8.5cm]{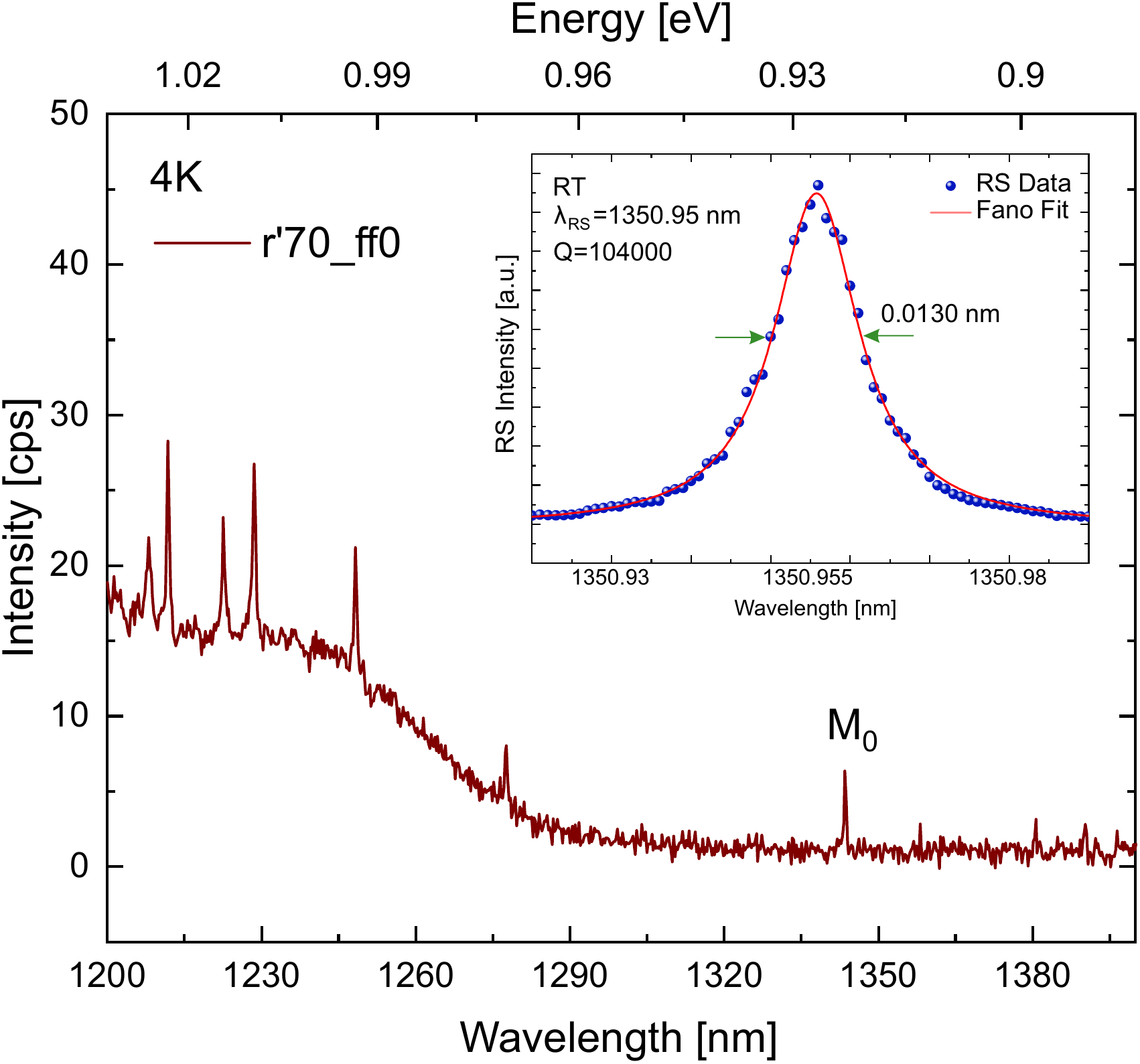}
\caption{Resonantly enhanced PL emission spectrum of the SQD centered in a N24 bichromatic PhCR with parameters a360 r70 without far-field optimization (ff0) as measured at 4K. The inset shows the resonant scattering spectrum of the same cavity measured at room temperature (blue symbols). The data were fitted to the Fano function \cite{Galli2009} plotterd in red. From the fit, a resonance wavelength of 1350.95 nm, a FWHM of 13.0 pm, and thus, a Q-factor of  $\sim 104000$ were obtained for the $M_0$ resonance. 
 }
\label{fig:ff0_Q100K}
\end{figure}

The inset in Fig.\,\ref{fig:comp3samples} shows the spectral profile of the $M_0$ resonance as observed at room temperature in a cross-polarized resonant scattering (RS) experiment \cite{Galli2009}. A tunable laser with a spectral resolution better than 1\,pm provided by a fiber-based reference Fabry-Perot cavity \cite{SimbulaAP2017} was scanned through the resonance of the PhCR (blue dots). The peak position of the RS signal is in agreement with the emission maximum of the $M_0$ mode observed at $T=290$\,K. The RS signal is fitted to a Fano-resonance as described in Ref.\cite{Galli2009} (red line in inset) and a very high Q-factor close to 53500 is obtained. We want to emphasize, that this value is limited by the effects of out-coupling structure (ff3) included in the respective PhCR layout. Omitting this structure (ff0) while leaving all other parameters of the PhCR unchanged (a=360\,nm, r=70\,nm, N=24) results in an even larger Q-factor as shown in the inset of Fig.\,\ref{fig:ff0_Q100K} by the blue dots and the fitted Fano-resonance (red line). In this case, we observe $Q\sim 104,000$, which is the largest Q-factor so far reported for a SOI based PhCR loaded with a QD emitter \cite{Kuruma2020,Schatzl2017,2015_OE_Zeng,Schwagmann2012,Nomura2009,Hennessy2007}. We ascribe this huge improvement of the Q-factor as compared to previous works  \cite{Kuruma2020,Schatzl2017,2015_OE_Zeng,Schwagmann2012,Nomura2009,Hennessy2007} mainly to our novel pit-pattern layout, together with the perfectly site-controlled QD growth technique, that allows restricting the number of QDs per PhCR to exactly the one centered within the cavity. With this layout, we avoid Q-factor degradation as a consequence of secondary QDs in the cavity that possibly act as photon scatterers or absorbers. In addition, via the differences in the etch rates of Si and SiGe for the employed plasma etching recipe, secondary QDs degrade the etch homogeneity and induce additional structural deviations from the ideal PhC structure. Again, we observe the RS resonance shifted to a larger wavelength by $\sim 11$\,nm  due to the different temperatures for PL and RS experiments. It is interesting to note that for the ff0-design the $M_0$ resonance is observed at 5\,nm longer wavelength as compared to the ff3-design. This finding is in agreement with a slightly smaller effective index of refraction in the ff3-design due to its smaller Si volume as a consequence of the enlarged air holes forming the ff structure.

\begin{figure*}[t]
\includegraphics[width=13cm]{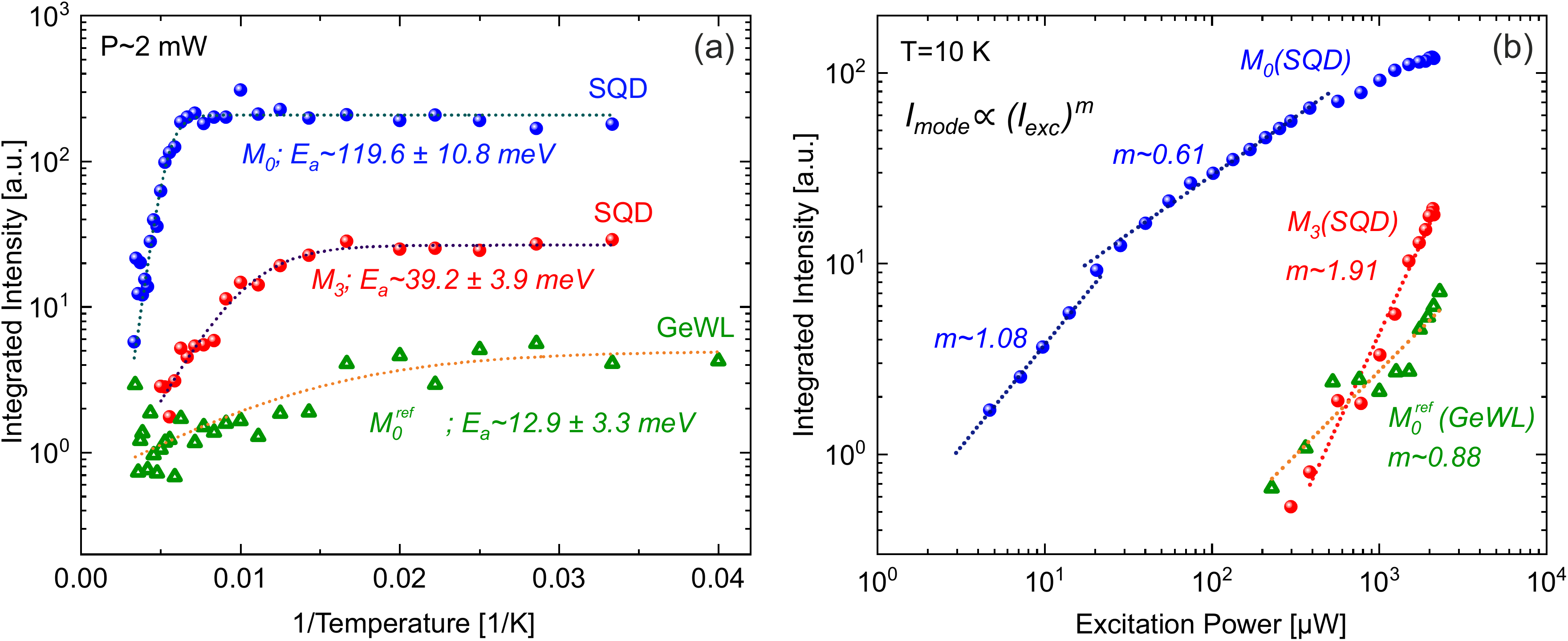}
\caption{(a) Temperature dependence of the 
 peak area of the three selected modes ($M_0$, $M_3$, and $M_0^\text{ref}$) as a function of reciprocal temperature. Arrhenius fits give thermal activation energies $E_a$ of each mode. (b) Peak area ($I_\text{mode}$) versus excitation power ($I_\text{exc}$) for modes $M_0$ and $M_3$ of the a360r70ff3 PhCR on the SQD sample (red an blue symbols, respectively). For mode $M_0^\text{ref}$ of the a350r70ff3 PhCR on the GeWL sample this dependence is shown by the green symbols. Dotted lines and $m$ values result from fits to $I_\text{mode} \propto (I_\text{exc})^m$ as described in the text. 
}
\label{fig:Tquench}
\end{figure*}

Further evidence for the distinctly different nature of the emission sources coupled to the modes $M_0$, $M_3$ and $M_0^\text{ref}$ of the PhCRs shown in Fig.\,\ref{fig:comp3samples} is provided by their different quenching behavior as the temperature rises from cryogenic to room temperature. Figure\,\ref{fig:Tquench}(a) shows Arrhenius plots of observed mode intensities integrated over the line shapes for modes $M_0$ (blue symbols), $M_3$ (red symbols) of the PhCR aligned to a SQD and for mode $M_0^\text{ref}$ (green symbols) of the PhCR on the WL sample. The data were fitted to the simplest quenching function given by \cite{1993_APL_Wachter}
\begin{equation}
    I(T)=\frac{I_0}{1+A\exp\left(-\frac{E_a}{kT}\right)}.
\end{equation}
Fitting results and extracted activation energies, $E_a$, are shown as dotted lines and labels in Fig.\,\ref{fig:Tquench}(a). The largest activation energy ($E_a=119.6\pm10.8$\,meV) is obtained for the $M_0$ resonance that we assigned to be fed by the SQD emission. Again we observe a distinctly different behavior for the $M_0^\text{ref}$ mode, for which we observe a more than 10 times reduced value for the activation energy of only $E_a=12.9\pm3.3$\,meV. The difference in the activation energies can be naturally understood by presence and absence of a QD in the PhCR's center on the SQD and WL sample, respectively.

As shown in Fig.\,S3 of the Supplementary Material, for the $M_0$ mode intensity of the a360r70ff0 design, we observe an activation energy identically within experimental errors to the one assigned to the QD emission in the previous paragraph. Thus, we have strong evidence that a QD is present also in the center of the a360r70ff0 PhCR. Nevertheless, only a small PL intensity is detected for the $M_0$ mode of this PhCR as shown in the main panel of Fig.\,\ref{fig:ff0_Q100K}, as a consequence of a reduced coupling to radiating modes for to the ff0 design. In turn, this reduced coupling manifests itself also in the large Q-factor for this design discussed in a previous paragraph.    

For the WL emission into mode $M_3$ of the PhCR on the SQD sample occurring at $\sim 1220$\,nm in Fig.\,\ref{fig:comp3samples}, we observe an activation energy ot $E_a=39.2\pm3.9$\,meV,  intermediate to the values reported for modes $M_0$ and $M_0^\text{ref}$ in a previous paragraph. In the same spectral range, very similar $E_a$-values are obtained for the emission peaks of the WL reference sample, as shown in Fig.\,S4 of the Supplementary Material.   

Also measurements of the emission intensity in the various PhCR modes $I_\text{mode}$ as  functions of the excitation intensity $I_\text{exc}$ are in agreement with our source assignments. In the double logarithmic plot shown in Fig.\,\ref{fig:Tquench}(b), individual results for modes $M_0$, $M_3$ and $M_0^\text{ref}$ as labelled in Fig.\,\ref{fig:comp3samples} are shown by blue, red and green symbols, respectively. For a characterization, we use the exponent $m$ of the commonly employed empirical relation $I_\text{mode} \propto (I_\text{exc})^m $. As shown in Fig.\,\ref{fig:Tquench}(b), for $I_{M_0}$  we obtain $m\approx 1$ for low and $m\approx 2/3$ for intermediate excitation power, with a transition region between 30 and 70\,$\mu$W. Such a behaviour is typically observed for QD emitters, with the $m=1$ region indicating dominant electron (e) hole (h) pair recombination and the $m=2/3$ region Auger-recombination involving three particles (for example two e and one h) \cite{Apetz1995, JulsgaardN2011}. At even higher excitation, additional e-h recombination paths result in a further decrease of $m$. On the other hand, for the $M_3$ mode we obtain $m\approx 1.9$, typically for WL emission \cite{Apetz1995}, where the carrier lifetime is limited by Shockley-Read- Hall recombination over trap states. For this recombination process, $m=2$ is expected \cite{2012_PP_Shklyaev}. Due to e, h localization in QDs, the trap states do not influence the carrier lifetime in QDs as long as they are not close to a QD. Thus, they are much more efficient for extended QW states. For mode $M_0^\text{ref}$ of the a350r70ff3 PhCR on the WL reference sample, we observe $m=0.88$, indicating e-h pair as the dominant recombination channel in this spectral region of the WL emission.

\begin{figure}[t]
\includegraphics[width=8cm]{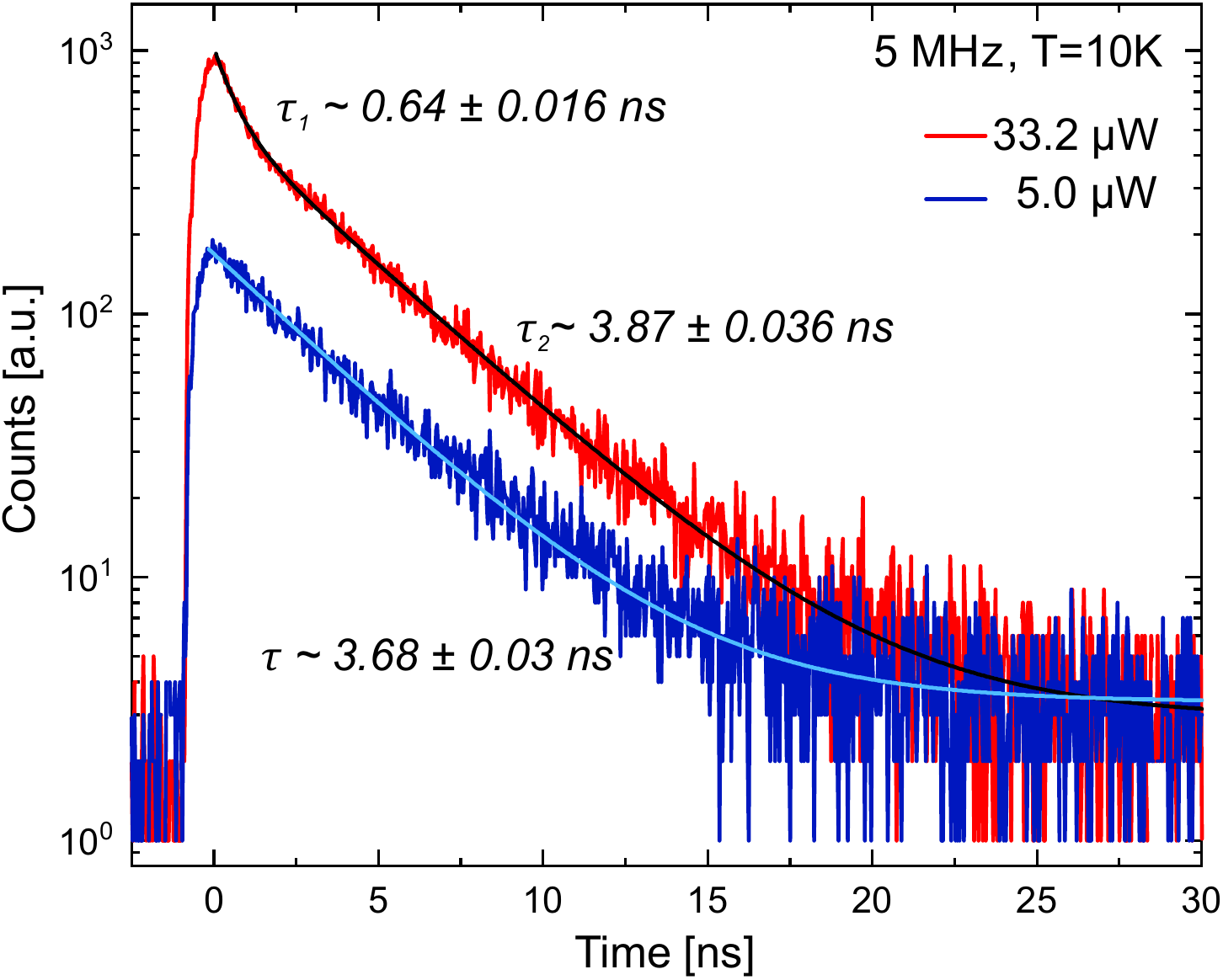}
\caption{Semi-logarithmic plot of time-resolved $M_0$ mode emission of a PhCR with parameters a360 r70 ff3 measured at 10K and recorded under 5 MHz repetition rate with 130\,ps laser pulses at 442\,nm. At 33.2\,$\mu$W average excitation power (red line) clearly a fast and a slow decay component are observed. At lower power (5\,$\mu$W, blue line) the fast component is absent.}
\label{fig:time_res}
\end{figure}

The decay dynamics of the $M_0$ emission was characterized by TCSPC experiments. The results of two measurements at different average laser power are shown in Fig.\,\ref{fig:time_res}. For an average excitation power of 33.2\,$\mu$W, an initial fast decay component with a decay time of $\tau \approx 0.65$\,ns is observed, followed by a slower decay with  a time constant $\tau \approx 3.9$\,ns. For a lower average excitation power of 5\,$\mu$W, a single exponential decay with a  time constant $\sim 3.7$\,ns close to the slow component at larger excitation is measured. According to Ref.\cite{JulsgaardN2011}, the observed decay characteristics is expected for QDs excited to more than one e-h pair. In this case, more combinatorial recombination paths for e-h pair recombination as well as Auger processes become effective, resulting in a reduced life time of the e-h density in the QD. Eventually, after the rapid initial decay only one e-h pair will be present in the QD, resembling the situation after weak excitation. Therefore, the slowest decay constants are expected to be equal for intense and weak excitation, in agreement with the experimental results shown in Fig.\,\ref{fig:time_res}.

\section*{Conclusion}
In this work we used CMOS compatible lithographic alignment to position single SiGe QDs at the electric field maxima of bichromatic PhCRs'  ground modes. Due to the high reproducibility of the alignment process, we were able to produce a large set of resonators with varying layouts, each one centered with respect to its single QD. By this lithographic tuning we could identify the cavity layout best matching the emission maximum of the coupled QD. The bichromatic cavity design provides very large Q-factors but requires extremely accurate alignment of QD and cavity, due to its narrow Si region at the cavity center. By reducing the amount of Ge within the resonator and the surrounding photonic crystal to a minimum of exactly one QD on its associated  WL, we improve the resonators' structural quality by improving the etching homogeneity for the fabrication process. In addition, by these measures we avoid light scattering out of the resonator due to secondary QDs. As a consequence, we were able to demonstrate a Q-factor  in excess of $10^5$ for a QD-coupled resonator mode. This value represents the largest Q-factor reported so far for QD-loaded photonic crystal resonators realized in a SOI integrated optics platform \cite{Kuruma2020,Schatzl2017,2015_OE_Zeng,Schwagmann2012,Nomura2009,Hennessy2007}. The dependence of mode emission intensities  on temperature, excitation power, and time after pulsed excitation was carefully analyzed  and compared to results from cavities fabricated on samples containing a WL without QDs. By this analysis we were able to identify PhCRs for which the emission coupled to the high Q ground mode  is associated with optical transitions within the SiGe QD. As these transitions occur in the relevant telecom wavelength band, the nano-optical system investigated in this work is of high relevance as a potential quantum optical source for a future SOI based integrated quantum photonic platform compatible with existing fiber networks. With respect to the latter aspect, the results of time resolved PL decay indicate that photon emission at a level of a single e-h pair is within reach. 


\bibliography{Optica_SiGe}

\clearpage
\section*{Methods}

\subsection*{Lithographic sample patterning} Sample patterning was performed by electron beam (e-beam) lithography using a Raith eline plus system. Prior to spin-coating the sample surface with a positive polymethyl methacrylate (PMMA) resist purchased from Allresist (Allresist 679.04), it was cleaned by dipping it into Acetone, Methanol and 10\% hydrofluoric (HF) acid for 30 s each. The HF dip promotes the adhesion of the e-beam resist. After exposure, the resist was developed for 60 s. The Raith eline plus system allows automatised alignment of a mask layer to reference marks by a pattern recognition algorithm.  \\
\subsection*{Pattern transfer} The pattern of the lithographic mask was transferred to top Si layer of the sample by reactive ion etching in an inductively coupled plasma setup (ICP-RIE) from Oxford Instruments (Plasmalab 100). To minimize lateral etching, we used a cryo-process, for which the sample was cooled to 183\,K by liquid $N_2$ cooling. A mixture of SF$_6$ and O$_2$ at flow rates of 5 sccm and 3 sccm, respectively, and an etch-chamber pressure of 15 mTorr was used as etch medium. The etch depths were verified \textit{ex-situ} by surface scans using a Digital Instruments Veeco Dimension 3100 atomic force microscope with a Nanoscope IV controller.

\subsection*{MBE growth} After cleaning the substrates in piranha solution \cite{Caro_1898} as well as  by a RCA  sequence,\cite{Kern_1970} and finally by a dip in 5\%- hydrofluoric acid (HF) to remove the native oxide, 
they were transferred into a solid-source MBE (Riber Siva 45) chamber for site-controlled SiGe QD growth in the etched pits \cite{ZhongAPL2003,Grydlik2013}. 
As first step, an \textit{in-situ} degassing at 650$^\circ$C for 15 minutes was performed. Then,  a 34\,nm thick Si buffer was grown at a growth rate of 0.7\,\AA/s while ramping  the substrate temperature from 450$^\circ$C  to 550$^\circ$C. The buffer layer buries potential surface defects and remaining contaminations and flattens the pit sidewalls by exposing $\{1,1,n>7\}$ low-surface-energy Si facets that enable site-controlled Ge QD nucleation \cite{Grydlik2013} at the pit center. The growth methods for realizing perfectly ordered SiGe QDs in very sparse regular arrays are central to this work. Thus, they are described in the main text. If required, a $\sim 115$\,nm thick Si cap was grown as final layer. As indicated in the inset of Fig.~\ref{fig:uncaped_perfect}(d), during the growth of the first $\sim38$\,nm of this cap the substrate temperature was ramped up from 500-700$^\circ$C to minimize Si/Ge intermixing \cite{Brehm2011}. The elevated growth temperature in the late state of the ramp is beneficial to achieving low defect densities in the Si capping layer. The remaining 77\,nm  capping layer were then grown at 700$^\circ$C as well. 

\subsection*{Optical characterization} The emission properties of the cavity coupled QDs were studied in a  micro-photoluminescence ($\mu$-PL) setup described in details elsewhere \cite{2019_AdP_Hackl}. For optical excitation, a continuous-wave (CW) diode laser emitting at 442\,nm wavelength (PicoQuant LDH-D-C-440) was focused via a microscope objective with numerical aperture NA=0.7 (Mitutoyo M Plan Apo NIR HR 100$\times$) to a PhCR. The sample was mounted on the cold finger of a liquid He cryostat (CryoVac Konti Mikro X/Y), allowing to cool the sample to less than 10\,K. Via x-y translation stages mounted in the isolation vacuum chamber of the cryostat, cold finger and sample can be positioned with $\sim 100$\,nm accuracy over $\sim 1$\,cm traveling range. The PL emission was collected  through the same objective and coupled to either a single- or a multi-mode fiber after filtering out the excitation laser. The multi-mode fiber was connected to an Acton SpectraPro 300i grating spectrometer with 300\,mm focal length, NA=0.25, equipped with 300 or 600  grooves/mm gratings at blaze wavelengths of 1 or 1.6\,$\mu$m, respectively. We detect the dispersed signal with a liquid-N\textsubscript{2}-cooled InGaAs line detector (Princeton Instruments OMA V), containing 1024 pixels at 25\,$\mu$m pitch. A single-mode fiber and superconducting single photon detector (Single Quantum ) operated in time-correlated photon counting (TCSPC) mode using a PicoHarp 300 time tagging electronics were used for time-resolved PL decay experiments.

\section*{Data availability}
 The main data supporting the findings of this study are available within the
article and its Supplementary Information. Extra data are available from the
corresponding authors upon reasonable request.

\section*{Acknowledgements }
 The authors acknowledge Alma Halilovic and Stephan Br\"{a}uer for cleanroom and other technical supports, as well as Lucio C. Andreani and Jeffrey Schuster for fruitful discussions. This work was supported by the project CUSPIDOR that has received funding from the QuantERA ERA-NET Cofund in Quantum Technologies implemented within the European Union's Horizon 2020, cofunded by the Italian Ministry of University and Research (MUR) and the Austrian Science Foundation FWF under Project I 3760-N27. Additional funding by the FWF under Projects FWF\_30564NBL (co-founded by the province of Upper Austria), Y1238-N36 and by the Linz Institute of Technology (LIT): Grant No. LIT-2019-7-SEE-114 is acknowledged. The Department of Physics of the University of Pavia is supported from MUR through the  “Dipartimenti di Eccellenza Program (2018-2022)”.  

\section*{Author contributions}

\noindent T.P., M.C., D.G., M.G., M.B. and T.F. designed the experiment. T.P., L.S., M.C., and M.G. carried out the experiment. J.A., L.V., F.S. and M.B. optimized the QD growth and provided the samples. F.F. and J.-M. H. adapted the SOI substrates. T.P. and T.F. analyzed the raw data and wrote the manuscript. All authors interpreted and discussed the paper's content and contributed to the final manuscript. T.F. supervised the project.

\section*{Additional Information }
\begin{itemize}
\item The authors declare no competing interests.

\item Supplementary material is provided on the following page. 
 
\item Correspondence and request for materials should be addressed to Thanavorn Poempool or Thomas Fromherz

\end{itemize}

\end{document}


\title{Supplementary Information: \\Single SiGe Quantum Dot Emission Deterministically Enhanced in a  High-Q Photonic Crystal Resonator}

\author{Thanavorn Poempool}
\email{thanavorn.poempool@jku.at}
\author{Johannes Aberl}
\affiliation{Institute of Semiconductor and Solid State Physics, Johannes Kepler University Linz, Altenbergerstra\ss e 69, 4040 Linz, Austria}
\author{Marco Clementi}

\altaffiliation[Current address: ]{Photonic Systems Laboratory (PHOSL), École Polytechnique Fédérale de Lausanne, Station 11, 1015 Lausanne, Switzerland}
\affiliation{Dipartimento di Fisica, Universit\`a di Pavia, Via Bassi 6, 27100 Pavia, Italy}
\author{Lukas Spindlberger}
\author{Lada Vuku\v{s}i\'c} 
\affiliation{Institute of Semiconductor and Solid State Physics, Johannes Kepler University Linz, Altenbergerstra\ss e 69, 4040 Linz, Austria}
\author{Matteo Galli} 
\author{Dario Gerace}
\affiliation{Dipartimento di Fisica, Universit\`a di Pavia, Via Bassi 6, 27100 Pavia, Italy}
\author{Frank Fournel}  
\author{Jean-Michel Hartmann}
\affiliation{University Grenoble Alpes, CEA, LETI, Grenoble, France}
\author{Friedrich Schäffler}
\author{Moritz Brehm}
\author{Thomas Fromherz}
\email{thomas.fromherz@jku.at}
\affiliation{Institute of Semiconductor and Solid State Physics, Johannes Kepler University Linz, Altenbergerstra\ss e 69, 4040 Linz, Austria}
\maketitle
\section{Photoluminescence spectra of QD ensemble}
\begin{figure}[hhhhh!!!!!!]
\centering\includegraphics[width=9cm]{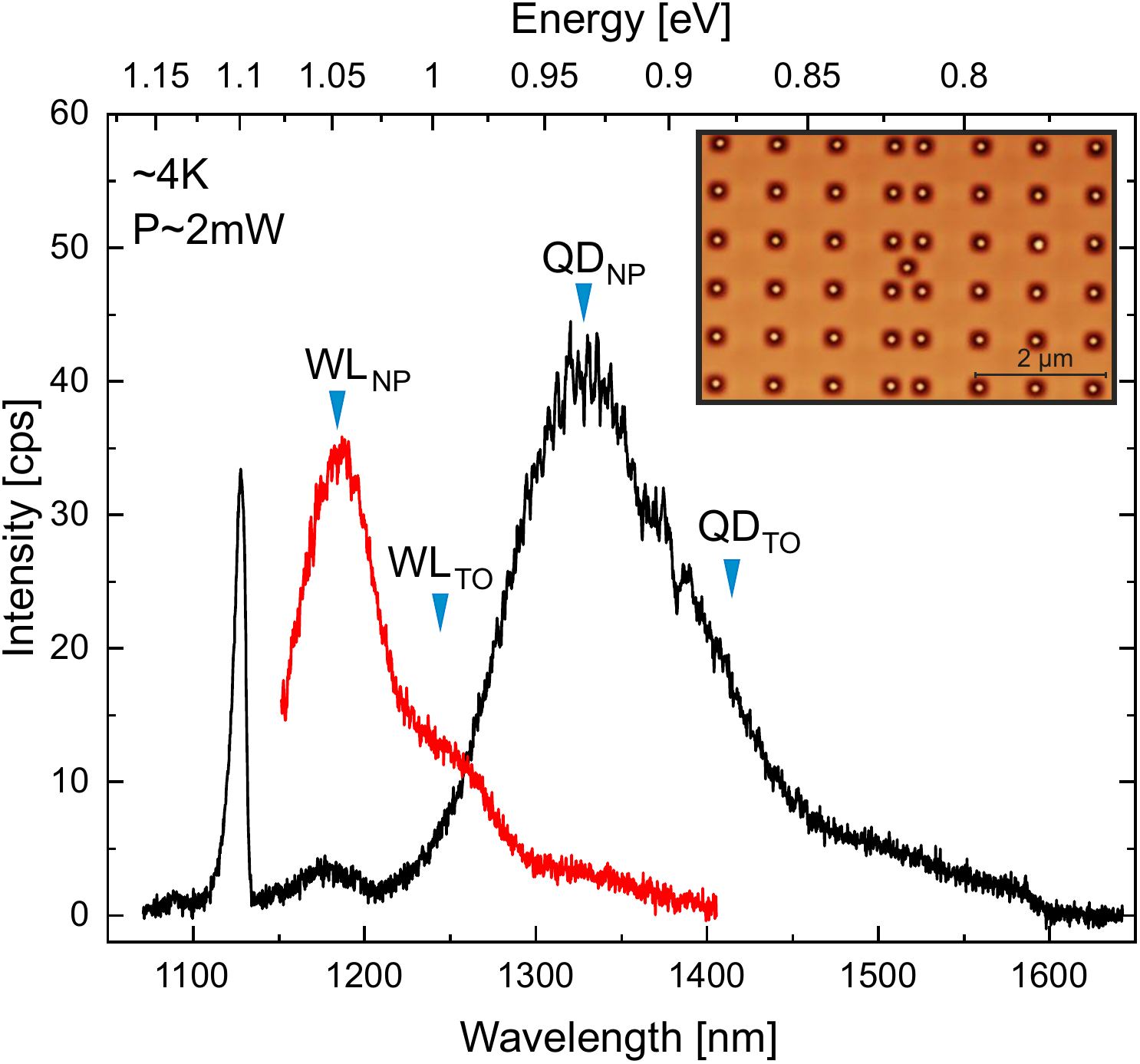}
\caption{Low temperature PL emission spectra from samples without PhCR. The black line shows the ensemble PL for densely ($\sim1.5\times10^8$\,cm$^{-2}$) ordered QDs, the red line the one of an unpatterned sample area without QDs where the Ge forms a two dimensional wetting layer. The comparison of these spectra clearly allows to identify the regions of WL and QD emission. The labels WL$_\text{NP}$, WL$_\text{TO}$, QD$_\text{NP}$ and QD$_\text{TO}$ indicate the no phonon (NP) emission and the emission shifted by a transverse-optical phonon (TO) energy for the WL and QDs, respectively. The inset shows an AFM image of the QD array as measured on an uncapped reference sample.}
\label{fig:ensemble_and_WL_PL}
\end{figure}

Figure\,\ref{fig:ensemble_and_WL_PL} compares the ensemble PL emission of densely ($\sim1.5\times10^8$\,cm$^{-2}$) ordered QDs as observed on a sample without PhCR with the emission of the WL without QDs. Due to the large density of the QDs, their PL emission is clearly observed also in the absence of a PhCR. Thus, the full emission spectrum of the QDs without the amplification due to the PhCR resonances can be observed. By comparing this spectrum with the spectrum measured in sample regions without pit-patter (red  line in Fig.\,\ref{fig:ensemble_and_WL_PL}), and, thus, without QDs, we can clearly identify the spectral range between $\sim 1300$\,nm and $\sim 1450$\,nm as being dominated by QD emission.          

\section{Lithographic tuning of resonator-coupled QD emission}

\begin{figure}[hhhhh!]
\centering\includegraphics[width=9cm]{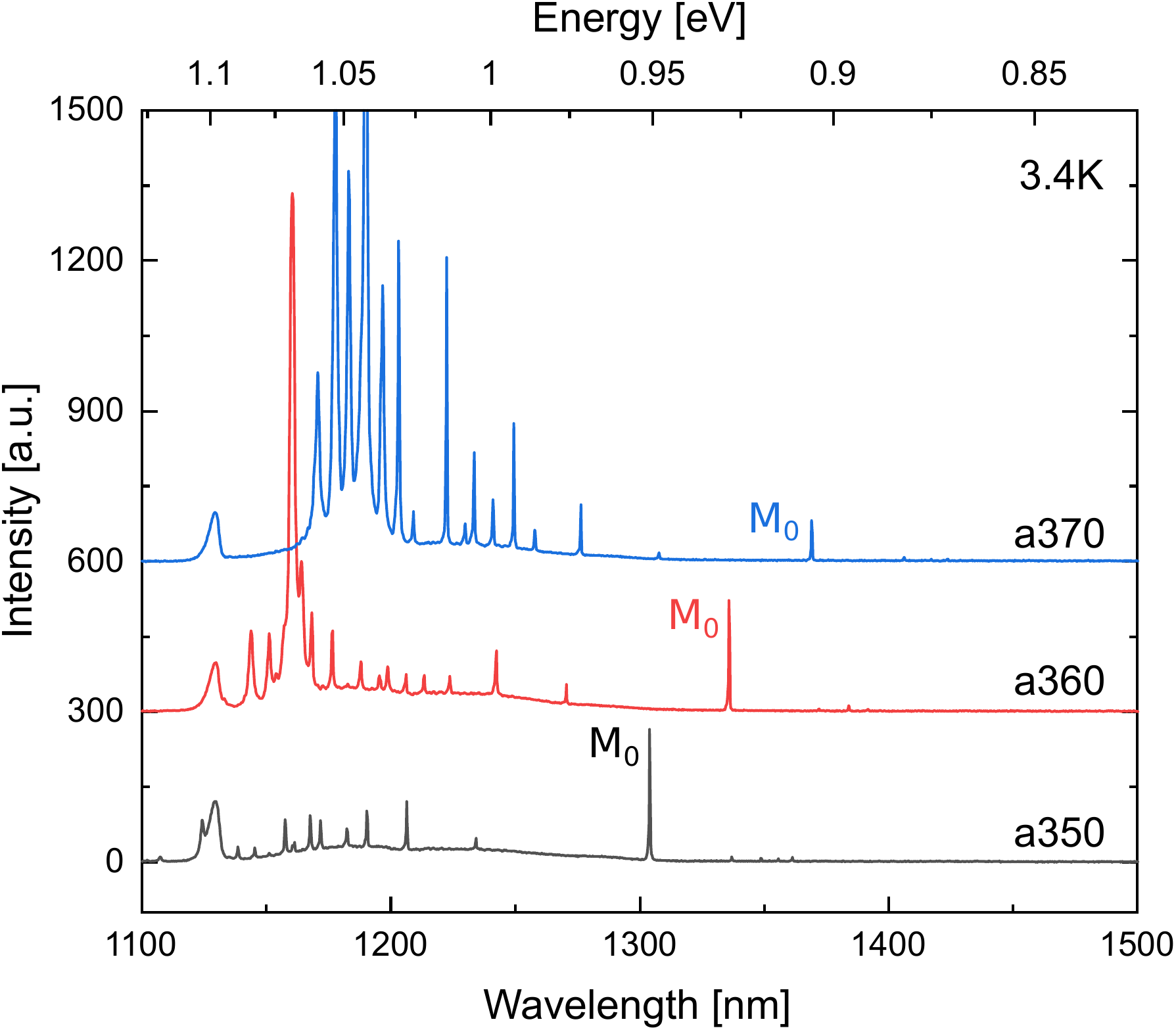}
\caption{PL spectra of single QDs coupled to bichromatic cavities with the different periods $a$ given by the line labels, but otherwise equal design parameters ($r=70$\,nm, ff=+3\,nm). The spectra were measured at the same excitation power (2.2\,mW) and are offset  vertically for clarity. The mode $M_0$ systematically shifts from $\sim\!1304$\,nm over $\sim\!1336$\,nm to $\sim\!1339$\,nm for $a=350,360,370$\,nm, respectively. For all $a$, the mode $M_0$ is comparably intense (within a factor 3), indicating a broad QD emission spectrum, in agreement with the results of Ref. \cite{Grydlik2015}.}
\label{fig:suppl_litho_tuning}
\end{figure}

Figure\,\ref{fig:suppl_litho_tuning} shows the PL emission of three bichromatic cavities with different periods $a$  but otherwise identical design parameters ($r=70$\,nm, ff=+3\,nm), excited with the same excitation power (2.2\,mW). The $M_0$ mode systematically shifts from $\sim1304$\,nm for $a=350$\,nm (black line) over $\sim1336$\,nm for $a=360$\,nm (red line) to $\sim1369$\,nm for $a=370$\,nm (blue line). For all these resonance wavelengths, we observe a variation of the $M_0$ mode intensity by approximately a factor 3. In the main paper, the $M_0$ emission was assigned to optical QD transition for the a360r70ff3 PhCR design. From this small variation, we  exclude a spectrally narrow emission spectrum of the QD in the center of the cavity, at least at the excitation intensities employed in this work. Our interpretation is in agreement with Ref.\,\cite{Grydlik2015}, where a broad PL emission of a single SiGe QD outside a PhCR is observed.  

\section{Activation energy of the $M_0$ resonance for a QD coupled PhCR without far-field optimization (ff0)}

\begin{figure}[hhhhh!]
\centering\includegraphics[width=8cm]{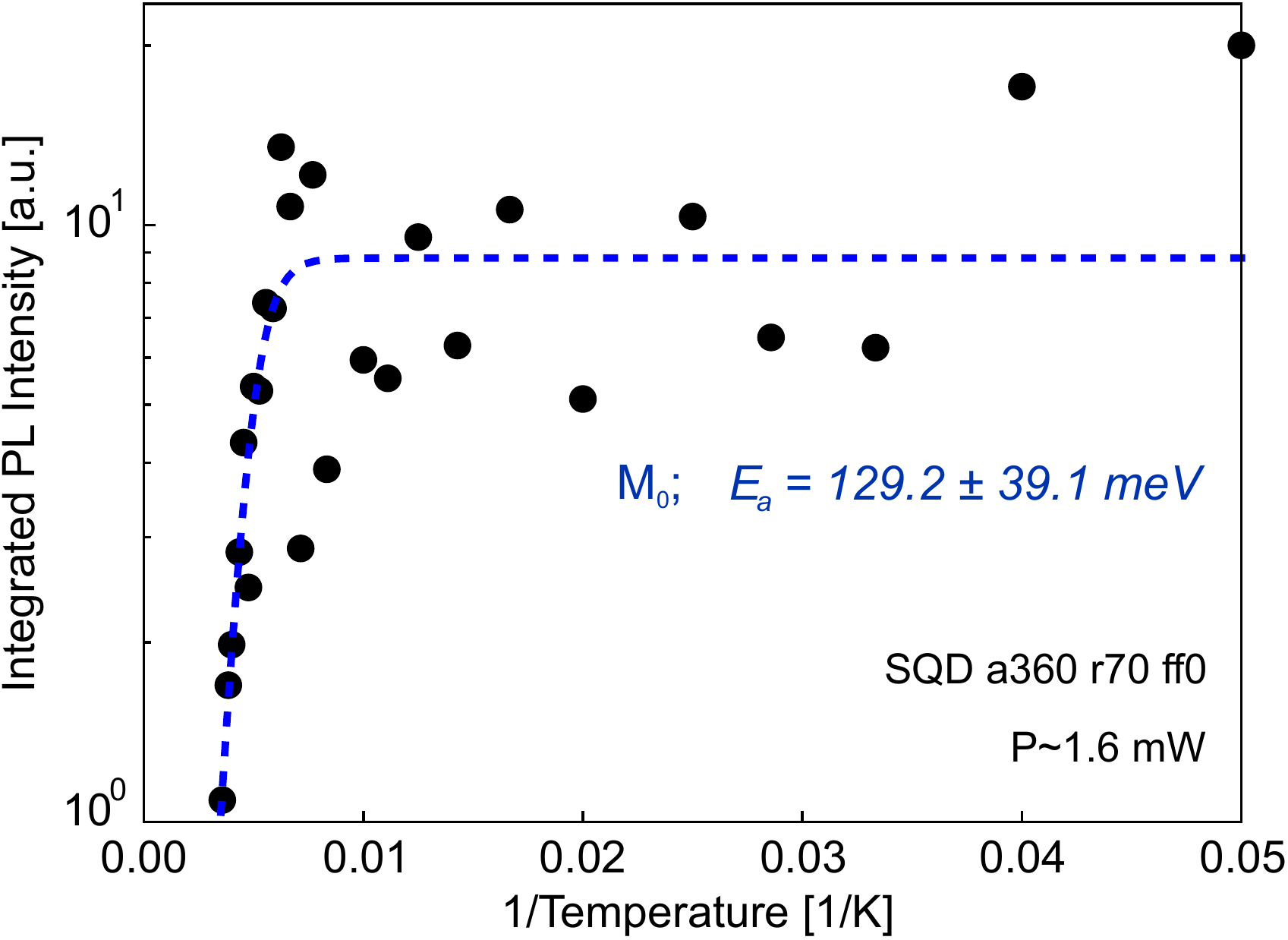}
\caption{Thermal quenching behavior for the $M_0$ mode PL intensity for a bicromatic PhCR without far-field optimization (a360r70ff0). The corresponding low-temperature PL spectrum is shown in 
Fig.\,3 
of the main paper. The activation energy $E_a$ as given in the plot is typical for optical QD transitions coupled to the PhCR and strongly indicates the presence of a QD in the PhCR center.}
\label{fig:suppl_Ea_M0_ff0}
\end{figure}

\begin{figure}[hhhhh!]
\centering\includegraphics[width=8cm]{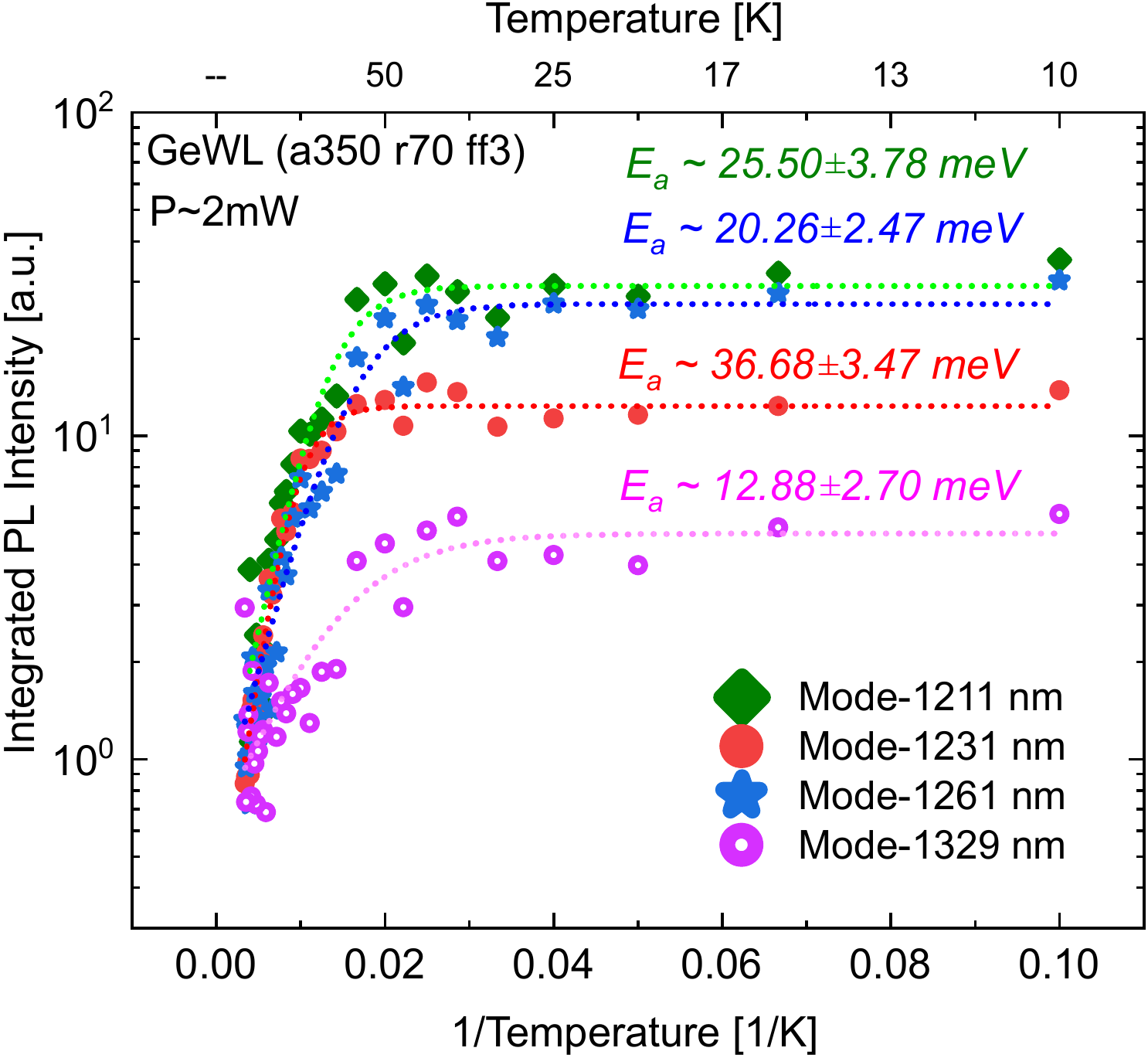}
\caption{Thermal quenching behavior of the mode intensities observed in PL for the a350r70ff3 bichromatic PhCR design fabricated on a WL reference sample. The corresponding PL emission spectrum is shown by the gray line in
Fig.\,2 
of the main paper. The different traces are labelled according to their resonance wavelength shown therein. Therefore, data shown in magenta correspond to the mode labeled $M_0^\text{ref}$ in 
Fig.\,2 
of the main paper. At each resonance wavelength, the fitted activation energies $E_a$ for WL PL quenching are much smaller than the one shown for the QD PL quenching in Fig.\,\ref{fig:suppl_Ea_M0_ff0} and in main-paper 
Fig.\,4 
by the blue line.}
\label{fig:suppl_Ea_WL}
\end{figure}

Due to the absent far-field optimization, the observed $M_0$ mode intensity shown in 
Fig.\,3 
of the main paper for a QD coupled to a a360r70ff0 PhCR is small. In fact, its intensity is comparable to the intensities observed for the $M_0^\text{ref}$ emission of the WL reference samples shown in Fig.\,2 
of the main paper. Thus, \textit{a priori} it can not be ruled out, that no QD nucleated in the pit and only a WL is present instead of the QD. In turn, the absent QD could possibly be the reason for the large Q-factor ($>10^5$) of the $M_0$ resonance reported for this PhCR design in the main paper. However, the thermal quenching behavior for the $M_0$ mode of this PhCR shown in the Arrhenius plot Fig.\,\ref{fig:suppl_Ea_M0_ff0} yields an activation energy $E_a=129.2\pm 39.1$\,meV, equal within the experimental error to the activation energy of a single QD coupled to the PhCR a360r70ff3 as found in the main paper. In addition, this activation is significantly larger than the activation energies obtained for any of the resonances shown in 
Fig.\,2 
of the main paper by the gray line for the WL reference sample. The corresponding Arrhenius plots are shown in Fig.\,\ref{fig:suppl_Ea_WL} and yield activation energies in the range between $\sim\!13$\,meV and $\sim\!37$\,meV for the four resonances labeled by their resonance wavelength in Fig.\,\ref{fig:suppl_Ea_WL}. Here the mode shown in magenta corresponds to mode labeled $M_0^\text{ref}$ in 
Fig.\,2 
of the main paper. These findings give strong evidence, that a QD is present in the center of the  a360r70ff0 PhCR and still Q-factors in excess of $10^5$ are feasible.

For completeness we want to add, that for extracting activation energies $E_a$, the logarithm of the observed mode intensities was fitted to the logarithm of the right-hand side of Eq. (1) given in the main paper. Such a fitting strategy is routinely employed in Arrhenius fits and emphasizes the significance of small experimental values that provide most information on a temperature induced quenching behavior.

\bibliography{Optica_SiGe}